\documentclass[a4paper]{article}
\usepackage{atmohead2013}
\usepackage[english]{babel}
\pdfoutput=1

\title{The Site of the ASTRI SST-2M Telescope Prototype:\\
Atmospheric Monitoring and Auxiliary Instrumentation}

\shorttitle{ASTRI SST-2M Site: Auxiliary Instrumentation}

\authors{
Giuseppe Leto$^{1}$,
Maria Concetta Maccarone$^{2}$,
Giancarlo Bellassai$^{1}$,
Pietro Bruno$^{1}$,
Mauro Fiorini$^{3}$,
Alessandro Grillo$^{1}$,
Eugenio Martinetti$^{1}$,
Giovanni La Rosa$^{2}$,
Alberto Segreto$^{2}$,
Giuseppe Sottile$^{2}$,
Luca Stringhetti$^{3}$,
for the ASTRI Collaboration$^{4}$.
}

\afiliations{
$^1$ INAF - Osservatorio Astrofisico di Catania, Via S. Sofia 78, I-95123 Catania, Italy \\
$^2$ INAF - IASF Palermo, Via U. La Malfa 153, I-90146 Palermo, Italy \\
$^3$ INAF - IASF Milano, Via E. Bassini 15, I-20133 Milano, Italy \\
$^4$ http://www.brera.inaf.it/astri\\
}

\email{Giuseppe.Leto@oact.inaf.it}

\abstract{ASTRI is a Flagship Project led by the Italian National Institute of Astrophysics, INAF.
The main objective of the ASTRI project is to develop a prototype of the Small Size class Telescope for the Cherenkov Telescope Array (CTA) in a dual-mirror configuration (SST-2M). The ASTRI SST-2M is an end-to-end prototype that will be fully developed by the ASTRI Collaboration from the optics design and manufacturing to the focal plane camera, from the structure of the mount to all the needed software.
The ASTRI SST-2M prototype will be placed at the INAF "M.G. Fracastoro" observing station in Serra La Nave on the Etna Mountain near Catania, Italy. The technological solutions adopted will be tested on field: observations of the Crab Nebula and of other sources will be essential part of the science verification phase, with the aim to assess the achievement of the scientific requirements.
In the following we present the Serra La Nave site together with all the auxiliary instruments needed for atmospheric monitoring and characterization, calibration and science verification of the ASTRI SST-2M prototype.
}


\begin{document}
\maketitle

%Begin a section.
\section{Introduction}
The study of astronomical sources in the range from tens of GeV up to 100 TeV and beyond, in the very
near future will take advantage of the forthcoming ground-based Cherenkov Telescope
Array, CTA \cite{bib:CTA_2011,bib:CTA_2013} that will explore the Very High Energy (VHE) domain with unprecedented
sensitivity, angular resolution and imaging quality.
These results will be achieved with the implementation of several tens of Imaging Atmospheric Cherenkov Telescopes
divided in three classes of configuration to cover and maximize the performance in the four energy decades of interest:
Large ($\sim24$m diameter), Medium ($\sim12$m diameter) and Small ($\sim4$m diameter) Size class Telescopes
are envisaged in the low, medium and high energy regime, respectively.
The CTA telescopes will be installed in two observatory sites, one for each hemisphere;
the Northern array will be mainly devoted to the study of
extragalactic sources while both the Galactic plane and the extragalactic
sky will be the main observation target for the Southern array.

The Italian contribution to CTA is mainly represented by the ASTRI Program \cite{bib:ASTRI_web}, a "Flagship Project" defined in 2011 and financed by the Italian Ministry of Education, University and Research. The ASTRI Program is led by INAF,
the Italian National Institute of Astrophysics, and the Collaboration includes about ten INAF Institutes and a few Italian Universities.

ASTRI is the acronym for "Astrofisica con Specchi a Tecnologia Replicante Italiana"; one of the items of the Program is the design, development and production of the mirrors for the CTA Medium Size Telescopes by using
an optimized version of an already well-tested replica technology \cite{bib:Canestrari_OE}.

The main and challenging item of the ASTRI Program is however the design, development and production,
within the CTA framework and following its requirements, of an end-to-end prototype of a Small Size class Telescope (SST)
devoted, with its full field of view of about 10 degrees, to the highest gamma-ray energy region ($>1$ TeV).
The prototype, named ASTRI SST-2M, will take advantage of many of the most innovative technological solutions present today and for the first time adopted all together in a Cherenkov telescope:
optical system in Schwarschild-Couder dual-mirror (2M) configuration \cite{bib:Pareschi_ICRC2013},
multi-pixels Silicon Photo Multipliers as sensors of the camera at the focal plane, readable via an
innovative and fast front-end electronics \cite{bib:Catalano_ICRC2013}.
A further objective of the ASTRI Program is the definition, development and deployment, at the CTA Southern site,
of an SST-2M mini-array which would constitute the first seed of the CTA Observatory.
For that reason the ASTRI SST-2M prototype design already takes into account all the hardware and software
prerequisites that are necessary in view of the future SST-2M mini-array.

The verification and scientific calibration phase of the ASTRI SST-2M prototype 
will be performed in Italy. The telescope will be placed at the INAF
"M.G. Fracastoro" observing station located in Serra La
Nave on the Etna Mountain near Catania, Italy. The installation
is foreseen in autumn 2014, immediately followed by
the start of the data acquisition. Here we briefly present the Serra
La Nave site, giving more emphasis to the complex of auxiliary instrumentation
that will be used on site to support the calibration and science
verification phase as well as the regular data reconstruction
and analysis of the ASTRI SST-2M prototype.

\section{The Site at Serra La Nave}
 \begin{figure*}[!t]
  \centering
  \includegraphics[width=0.9\textwidth]{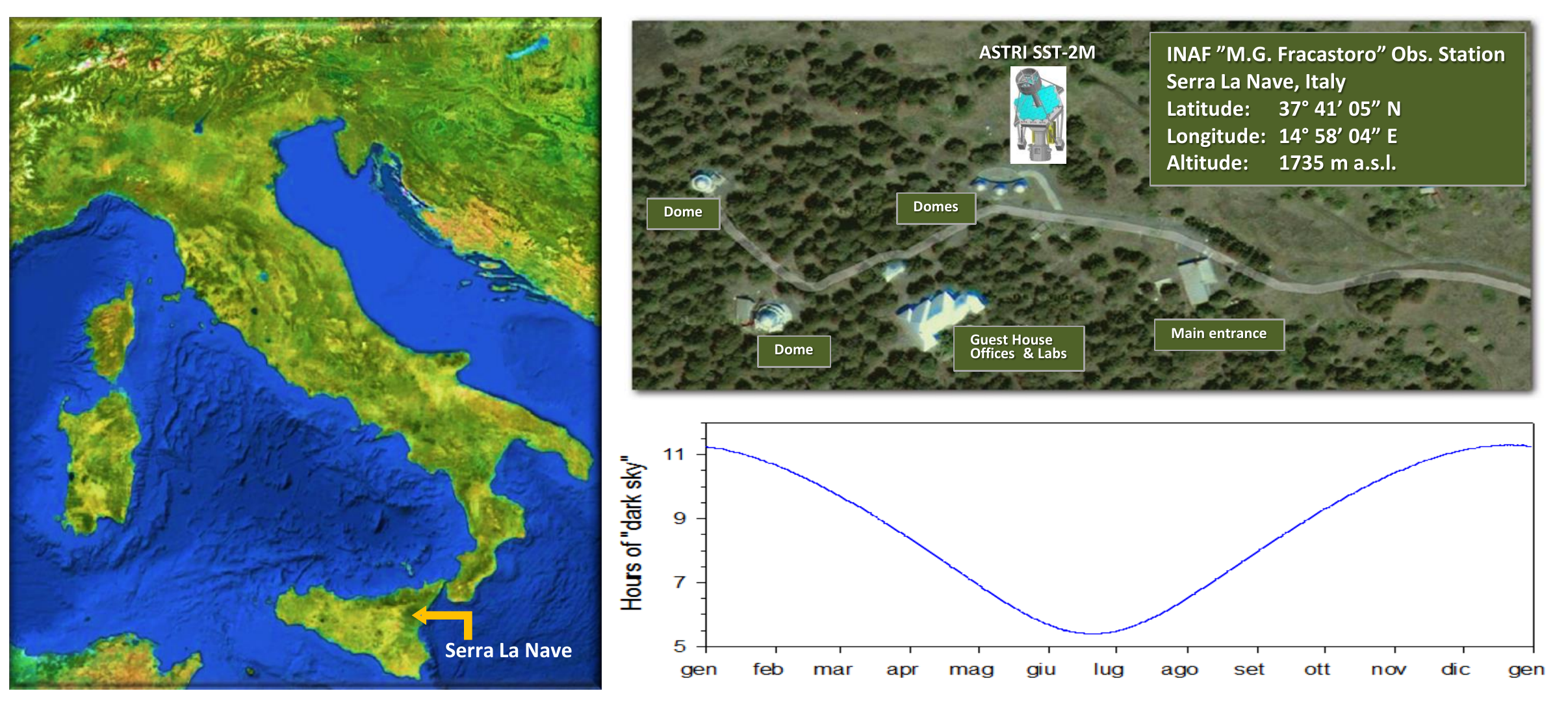}
  \caption{Location of Serra La Nave (left) and aerial view of the "M.G. Fracastoro" observing station (upper right panel); superimposed is the ASTRI SST-2M sketch. Lower right panel: the time extension of "dark nights" at Serra La Nave.}
  \label{SLN_location}
 \end{figure*}

The ASTRI SST-2M prototype will be tested on field in order to verify its overall ability to fulfill the science objectives that CTA collaboration has indicated. The site where install ASTRI SST-2M telescope must obviously satisfy some general geophysical conditions needed to detect VHE gamma-rays using the Cherenkov observational approach; moreover, the site must satisfy requirements strictly related to atmospheric and meteorological conditions as well as accessibility and infrastructures. For this purpose, we performed a detailed review of the INAF candidate observing stations present in the National territory \cite{bib:Maccarone_TN_Site}. The final choice was in favor of the Serra La Nave (SLN) site,
whose figures of merit are compliant with the general specification required for SST telescopes by the CTA Collaboration \cite{bib:CTA_2011}, obviously restricted to the case of a single telescope.

The Serra La Nave observing station is at 1735 m a.s.l, 37$^\circ$ 41' 05" N Latitude, 14$^\circ$ 58' 04" E Longitude. The station is inside the "Parco dell'Etna" on the southern side of the Etna Mountain (Figure \ref{SLN_location}), in a land protected from strong wind and from much of the fallout of volcanic ash which influences the transparency of the atmosphere. Although the region is defined with medium risk of seismicity, no earth-faults are present in territory so that the earthquake risk is strongly reduced \cite{bib:Leto_IR_seismic}. Tremors due to the Etna activity are modest; volcanic ash can be present, depending on the wind direction, but only in minor content and for only few days per year. Ultimately, the SLN observing station has been identified as the best INAF Italian site for the installation, calibration and scientific verification of the ASTRI SST-2M Cherenkov telescope.

From the observational point of view, the main features of SLN can be summarized as follows \cite{bib:Maccarone_ICRC2013}:\\
- horizon clear above 20$^\circ$ on average;\\
- atmosphere transparency inside the limits;\\
- volcanic ash rarely present;\\
- average humidity of $67\%$ (summer) and $79\%$ (winter);\\
- minimum temperature of $-10^\circ$C (winter, night);\\
- maximum temperature of $28^\circ$C (summer, day);\\
- average wind speed of 1.9 m/s (7 km/h);\\
- very rare wind gusts (78 km/h max);\\
- cloud coverage  $<50\%$;\\
- fraction of useful nights $>53\%$;\\
- time extension of "dark nights" \footnotemark  \footnotetext{dark night: time interval from the end of astronomical twilight in the evening to the beginning of astronomical twilight in the morning, when the Sun is under 18$^\circ$ below the horizon.} from about 6 hours (July) to more than 11 hours (January).

The light pollution is present at medium level extending in South-South-East direction under 30$^\circ$ of elevation. In order to have an up-to-date evaluation of the sky brightness, an observational campaign has been done just after New Moon in July 2011. Measurements in U, B, V bands were made with the robotic APT2 telescope at the SLN site in order to map the sky brightness in a evenly spaced grid. During observation the mean temperature was 13$^\circ$C with 63\% relative humidity.
The overall result was that a source close to the Crab Nebula in its apparent path as seen from the SLN site, is observed on a sky whose magnitude in U is less bright than 20.4; in the case of B, the magnitude will be greater than 20.7, while in the case of V is maintained near the 20.0 for a large part of the trajectory.

From the logistics and infrastructure point of view \cite{bib:Maccarone_ICRC2013}, the SLN observing station is located at about 30 km from the laboratories of the Catania Astrophysical Observatory (OACT),
where the ASTRI SST-2M camera will be assembled and characterized, and at a short
distance ($<2$ km) from inhabited areas and from safety/emergency
structures ($<15$ km).
The station is provided with Internet access via  optical
fiber and has WiFi connection within the area. The station, accessible all year long, is
equipped with some small apartments that will be available for the ASTRI
staff during the various phases of installation, testing and operation of
the SST-2M prototype.

\section{Atmosphere monitoring}
In the case of the Cherenkov observations, the atmosphere is the primary detector that is used to reveal high energy photons. The conditions of the atmosphere at the site during observations are vital in order to be able to accurately reconstruct the parent event from the data observed by a Cherenkov telescope. That is why there is a number of instruments, external to the telescope, needed to monitor atmospheric conditions of the site. We refers to those instruments as "auxiliary".

The main outcomes from the SLN auxiliary equipment concern weather and environmental information, including sky brightness and atmosphere attenuation.

\begin{figure}[t]
  \centering
  \includegraphics[width=0.49\textwidth]{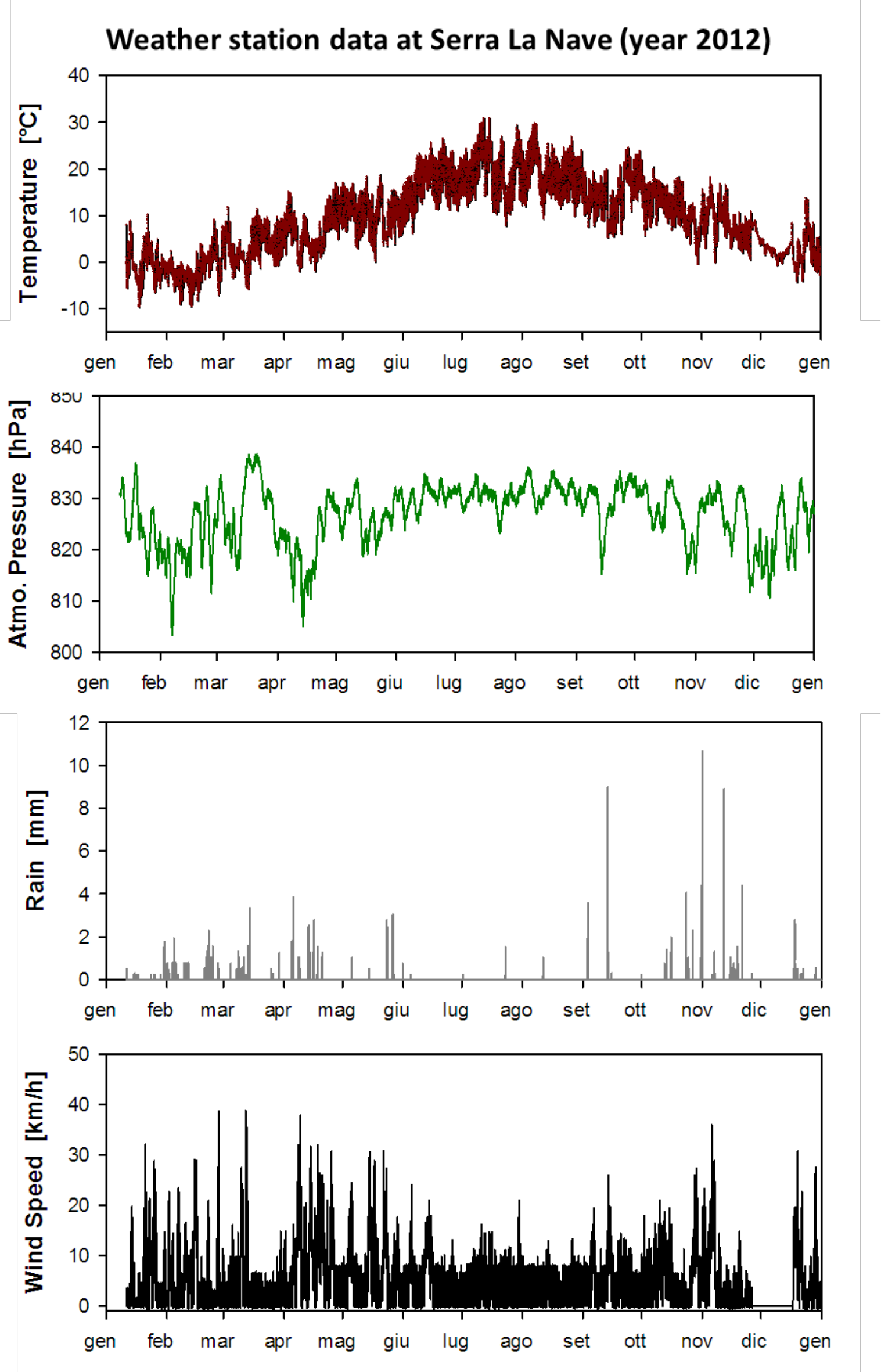}
  \caption{Profiles of temperature, pressure, rain and wind speed (averaged in 15 minutes time window) at Serra La Nave during the year 2012.}
  \label{weather_2012}
 \end{figure}

ASTRI SST-2M prototype, such as for all the CTA telescopes, will perform observations unless the weather does not allow them; this control is managed by the continuously active weather station, a Vantage Pro2 Davis Instrument \cite{bib:WSDavis} equipped with a Weather-Link Streaming Data Logger. Parameters like temperature, humidity, pressure, wind speed and direction, rain and rain rate will be logged and archived at a rate up to a entire set every 2 seconds. As an example, Figure \ref{weather_2012} shows data acquired at SLN during 2012. To improve accuracy and timeliness, a further rain sensor will be installed at SLN. The meteo station and rain sensor system will then provide alarms for the necessary parameters, commanding the activation of safe procedures of all the telescope functions whenever any alert is issued.

A further alarm can be provided from the lightning detector that will be installed at the SLN site in few months. The detector is an Electric Field Meter (EFM) \cite{bib:EFM} developed from a research group of the INAF Radio Astronomy Institute (IRA) in Bologna; the EFM will be used for monitoring atmospheric electric field variations and to issue alerts to the telescope control system. In an array, EFM is also capable to give detailed information about the direction of the lightening activity and the approaching speed.

The sky conditions in SLN will be primarily monitored by a fish-eye all-sky camera and a sky quality meter.

The all-sky camera is the SBIG AllSky-340C color fish-eye model \cite{bib:AllSkyCamera} which provides monitoring of cloud coverage both during daylight and night time, allowing a continuous monitoring of the cloudiness for statistical and forecast purposes. The all-sky camera is equipped with interfaces for the PC connection and data storage. Images can be saved at different time intervals; from the stored images a log of the cloudiness of the sky in the field of view (FoV) of the ASTRI SST-2M prototype will be created.

To have a prompt parameter for the evaluation of the sky quality during observations we use a Sky Quality Meter - LE (SQM) that measures the brightness of the night sky in magnitudes per square arcsecond with a 10\% precision ($±0.1$ mag/arcsec$^2$). The SQM is sensitive only to visual light and the model installed at SLN presents a Half Width Half Maximum of the angular sensitivity equal to ~10$^\circ$. The system returns integral information about background light intensity inside the FoV (~20$^\circ$) on demand up to a frequency of 1 Hz. The values of the measured sky brightness will be registered in the log file of the observation performed with the ASTRI SST-2M prototype.

As concerns the atmosphere quality monitoring to be performed during observations, at the SLN observation station there are two more instruments, PLUDIX and a Raman LIDAR, that are part of a collaboration of the OACT with the Istituto Nazionale di Geofisica e Vulcanologia, Osservatorio Etneo, Sezione di Catania (INGV-Catania).

\begin{figure}[t]
  \centering
  \includegraphics[width=0.4\textwidth]{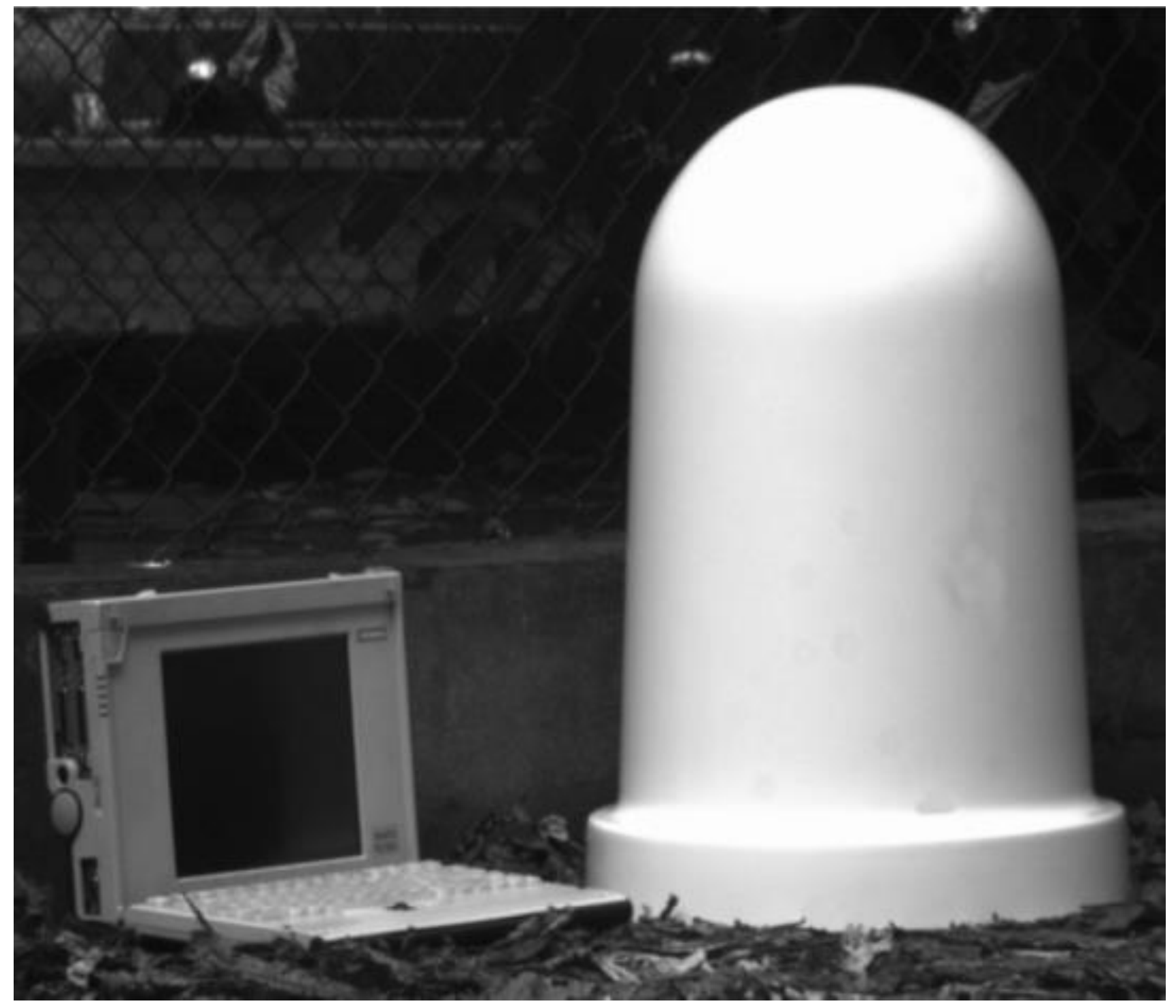}
  \caption{The PLUDIX sensor with its acquisition data equipment.}
  \label{pludix}
 \end{figure}

The PLUDIX is a Pluviodisdrometer consisting in a small radar that
exploits the Doppler effect of the particles in motion, detects the amount and the particle size and
provides an estimate of the quantity falls to the ground. PLUDIX is then a detector of the presence of precipitation and can identify the type of precipitation (rain, snow, hail, ash), measure the size distribution of hydrometeors (drops, snowflakes, hailstones), measure intensity of the instantaneous precipitation (rain gauge), reveal ash fall.
A model of PLUDIX is shown in Figure \ref{pludix}.

The LIDAR (LIght Detection And Ranging) allows to study the atmospheric composition,  structure, clouds and aerosol through the measurement of the atmospheric extinction profile. The LIDAR mounted at SLN can perform pointed observations and will be used to monitor atmospheric conditions up to 15 km height.

The auxiliary system at SLN currently foreseen is eventually completed by the UVscope \cite{bib:Maccarone_NIMA} and UVSiPM instruments \cite{bib:Sottile_SCINEGHE}, two portable multi-pixels photon detectors operating in single counting mode, entirely developed at the IASF-Palermo/INAF Institute.
Their acquisition data will support the monitoring of the night sky background and of the atmosphere attenuation, as well as the absolute calibration of the whole ASTRI SST-2M prototype system.

 \begin{figure}[t]
  \centering
  \includegraphics[width=0.4\textwidth]{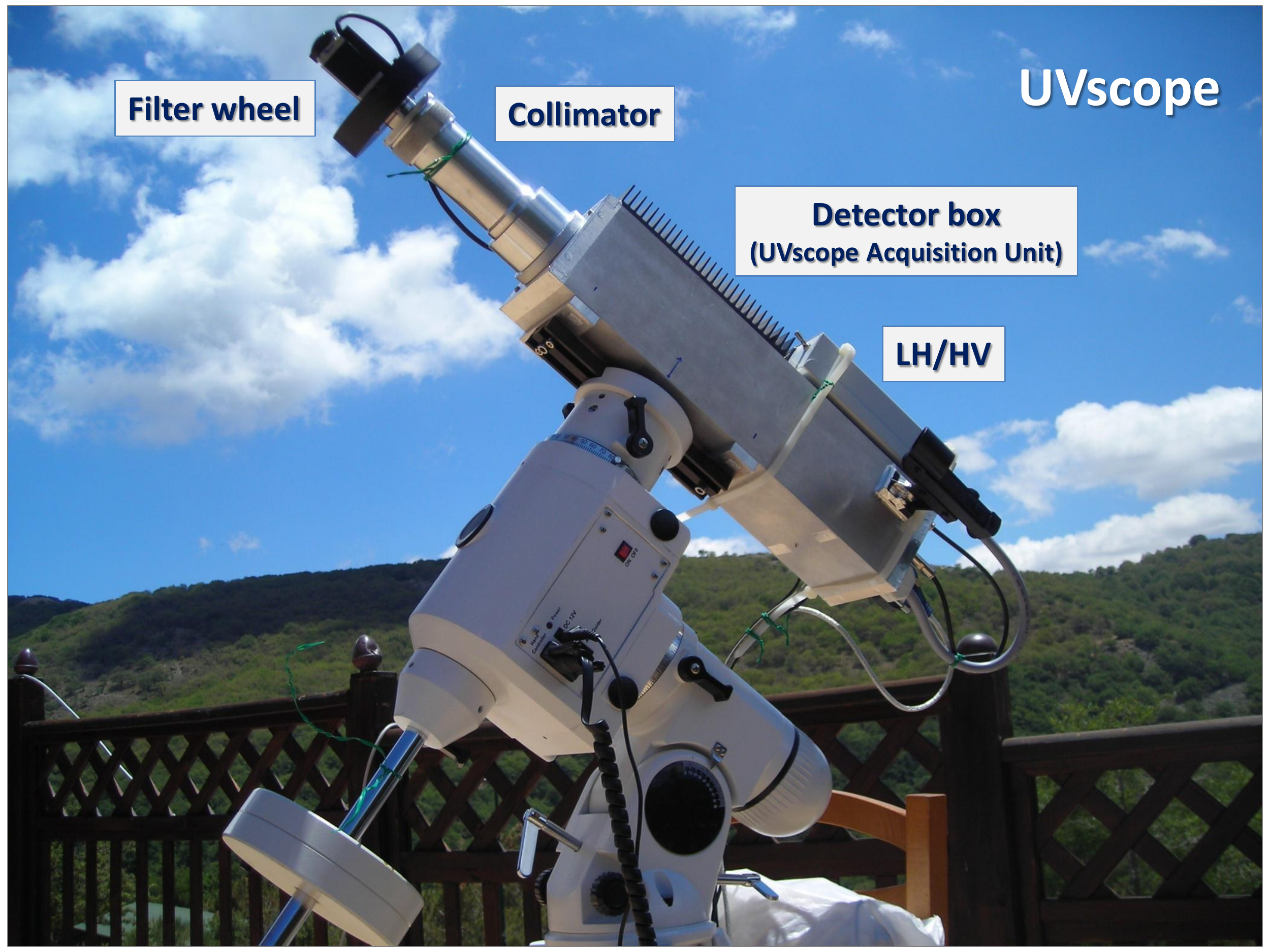}
  \caption{UVscope on its mount during the 2009 campaign in Contrada Pomieri, Italy (1335 m a.s.l.) devoted to the study of night sky background and atmospheric transparency at several wavelengths making use of a motorized filter wheel \cite{bib:Maccarone_NIMA}.}.
  \label{UVscope}
 \end{figure}

 \begin{figure}[t]
  \centering
  \includegraphics[width=0.4\textwidth]{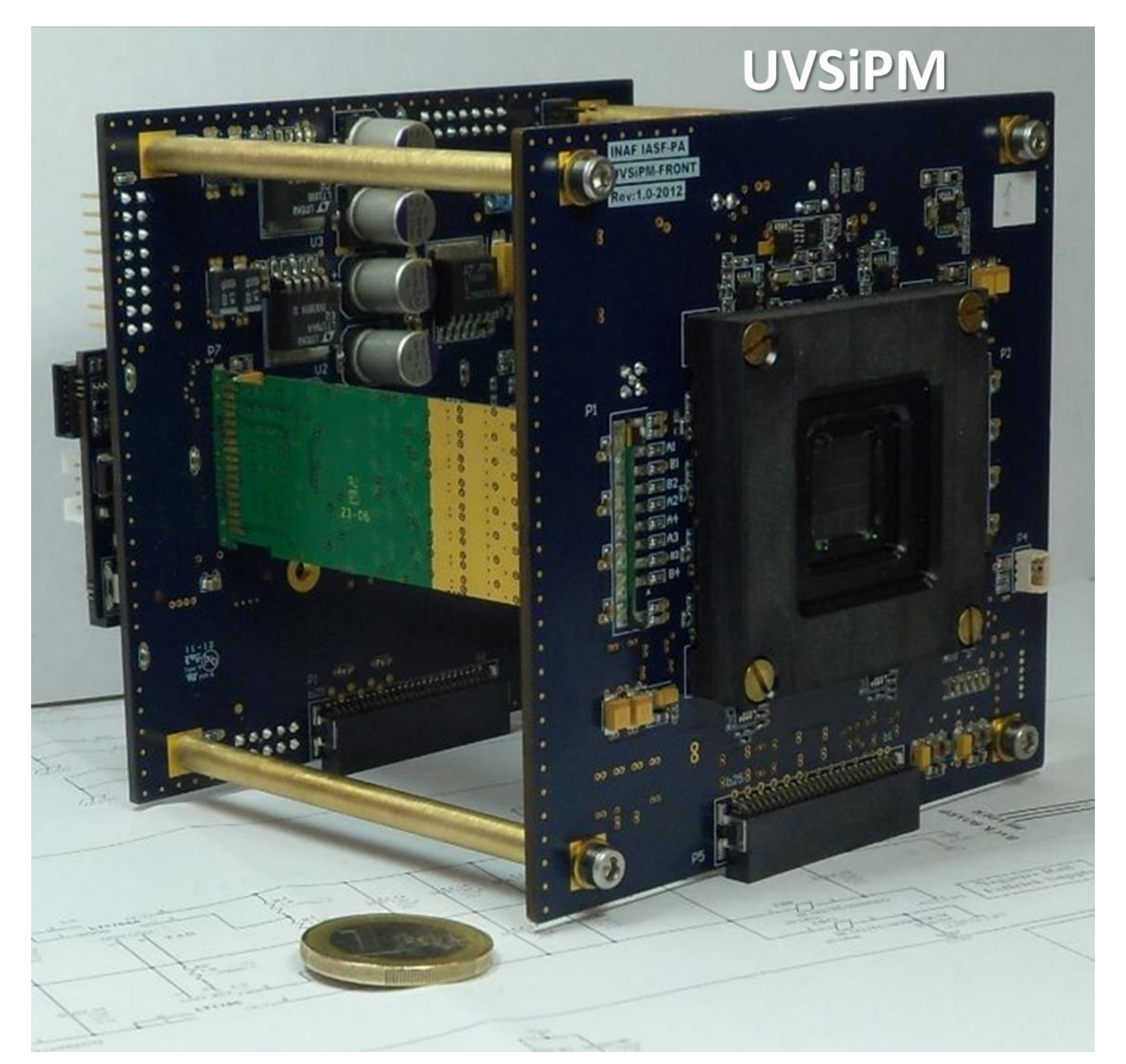}
  \caption{Inside UVSiPM: front view of the acquisition unit with the
socket for the SiPM sensor \cite{bib:Sottile_SCINEGHE}.}
  \label{UVSiPM}
 \end{figure}

The UVscope (Figure \ref{UVscope}) sensor is a multianode photo-multiplier (wavelengths range 300-650 nm), while the UVSiPM (Figure \ref{UVSiPM}) sensor is a Silicon photomultiplier (wavelengths range 320-900 nm) of the same model of sensors that will fill the camera at the focal plane of the ASTRI SST-2M prototype. The two detectors forming  UVscope-UVSiPM will be configured with proper entrance pupil and collimator length in order to obtain the same FoV; moreover, both the instruments will be completed with equal calibrated filters inside their collimators. Both  UVscope and UVSiPM, mounted on a motorized SmartStar MiniTower Pro, will be moved contemporarily pointing the same source direction without any interference with the ASTRI SST-2M operations. Several kinds of acquisitions are foreseen; as an example, UVscope-UVSiPM can measure the diffuse emission of the night sky background in the ASTRI SST-2M field of view, allowing a real time gain monitoring. During clear nights, both ASTRI SST-2M and UVscope-UVSiPM would simultaneously "track" a reference star pointing at the RA-Dec star position and following it at different elevation angles; thanks to the accurate calibration of UVscope-UVSiPM performed in lab, the flux profiles will allow us to determine the total atmospheric attenuation and the absolute calibration constants for the prototype. Last but not least, UVscope-UVSiPM will be used, simultaneously with ASTRI SST-2M, to observe a ground light source of well-known properties; the comparison of the separated but simultaneous acquisitions will result in the determination of the spectral response of the whole ASTRI SST-2M telescope, including both optics and camera \cite{bib:Segreto_CCF2013}.

\section{Conclusions}
The ASTRI SST-2M telescope, prototype of the Small Size class Telescope for the CTA, will be tested on field at the INAF observing station located in Serra La Nave, Italy. A set of different devices, external to the telescope, is required to perform its technological and scientific verification.

The complex of auxiliary instrumentation described in this document is already under installation at the SLN site and will support the calibration and science verification phase
as well as the regular data reconstruction and analysis of the ASTRI SST-2M prototype.\\

\vspace*{0.5cm}
\footnotesize{{\bf Acknowledgment:}{This work was partially supported by the ASTRI "Flagship Project" financed by the Italian Ministry of Education, University, and Research (MIUR) and led by INAF, the Italian National Institute of Astrophysics. We also acknowledge partial support by the MIUR 'Bando PRIN 2009'. INAF-OACT also acknowledges the support for atmospheric monitoring by PO Italia Malta 2007-2013 with VAMOS SEGURO project, A1.2.3-62 in the framework of the collaboration of OA Catania with the INGV-Catania .}}

\end{document}